%% file: main.tex
\documentclass[conference]{IEEEtran}

\ifCLASSINFOpdf
\else
\fi

\usepackage{float}
\usepackage{soul, color, acronym, multirow, amsmath, balance}
\usepackage[pdftex]{graphicx}
\usepackage[justification=centering]{caption}
\usepackage{tikz}

\input{acr.tex}

\newcommand\copyrighttext{%
  \footnotesize This work is copyrighted by the IEEE. Personal use of this material is permitted. However, permission to reprint/republish this material for advertising or promotional purposes or for creating new collective works for resale or redistribution to servers or lists, or to reuse any copyrighted component of this work in other works must be obtained from the IEEE. This work has been accepted for publication in IEEE ICC 2017, 21-25 May, Paris.}
\newcommand\copyrightnotice{%
\begin{tikzpicture}[remember picture,overlay]
\node[anchor=south,yshift=10pt] at (current page.south) {\fbox{\parbox{\dimexpr\textwidth-\fboxsep-\fboxrule\relax}{\copyrighttext}}};
\end{tikzpicture}%
}

\begin{document}

%
\title{Application Protocols enabling Internet of Remote Things via Random Access Satellite Channels}

\author{\IEEEauthorblockN{Manlio Bacco, Marco Colucci, Alberto Gotta}
\IEEEauthorblockA{Institute of Information Science and Technologies (ISTI)\\
National Research Council (CNR), via G. Moruzzi, 1, Pisa (Italy)\\
e-mails: \{name.surname\}@isti.cnr.it}
}

\maketitle

\begin{abstract}
Nowadays, \ac{M2M} and \ac{IoT} traffic rate is increasing at a fast pace. The use of satellites is expected to play a large role in delivering such a traffic. In this work, we investigate the use of two of the most common \ac{M2M}/\ac{IoT} protocols stacks on a satellite \ac{RA} channel, based on DVB-RCS2 standard. The metric under consideration is the completion time, in order to identify the protocol stack that can provide the best performance level.
\end{abstract}
\copyrightnotice 


%
\IEEEpeerreviewmaketitle

\section{Introduction}
\label{sec:introduction}
Satellite-based \ac{M2M} communications represent a large fraction of the \ac{IoT} market, showing an increasing popularity both in the research and in the industrial community. The ubiquitous coverage provided by the satellites may represent a key feature to enable the so-called \ac{IoT} massive internetworking, bringing the connectivity even in remote areas that are unlikely to be covered by other communication infrastructures (e.g., cellular).

In order to evaluate the feasibility of a satellite-based solution, along with any physical layer issues (e.g., \ac{SNR}, power consumption), the interactions between the transport and the application layer protocols need further investigations. 
Connection-oriented transport protocols, like \ac{TCP}, require connection establishment procedures, the use of flow control or congestion control algorithms, which may increase the communication overhead. The latter is an issue that needs to be carefully taken into account, especially in the case of \ac{IoT}/\ac{M2M} short-lived connections. In order to mitigate the aforementioned issue, \ac{IETF} has proposed the use of \ac{CoAP} 
(RFC 7252), a lightweight protocol designed for resource-constrained devices. It relies on the use of \ac{UDP} at the transport layer; because of this, the reliability is left out as an optional feature, to be implemented at the application layer. \ac{CoAP} endpoints exchange messages according to a request/response mode and the resources are accessed through a \ac{URI}.
In order to avoid a polling mechanism, \ac{IETF} has designed a protocol extension to \ac{CoAP}, based on the so-called \textit{observer} design pattern (RFC 7641). 
\ac{CoAP} clients \textit{register} to the \ac{CoAP} server; then, each client receives a \textit{notification} every time the state of a resource changes. The observer pattern is somewhat similar to the \ac{PUB/SUB} paradigm \cite{pubsub}, as implemented by MQTT, 
for instance. The performance provided by the use of MQTT on \ac{RA} satellite channels, according to the \ac{DVB-RCS2} standard \cite{dvbrcs2}, has been preliminary studied in \cite{advances}. Contrarily to \ac{CoAP}, MQTT is \ac{TCP}-based, thus the congestion control algorithm at the transport layer of each \ac{RCST} is responsible for rate control and retransmissions, if any erasures occur on the satellite channel.

In this work, we propose a comparison between the MQTT-based scenario in \cite{advances}, and a \ac{CoAP}-based protocol stack. The performance metric under consideration is the \textit{completion time}, which is the time a producer takes to successfully deliver data to a consumer.
The rest of this paper is organized as follows: Section \ref{sec:rlatedWorks} reviews some of the most relevant works in the literature, focusing on \ac{IoT}/\ac{M2M} communication scenarios via satellite and on the comparisons between \ac{IoT}/\ac{M2M} protocol stacks. Section \ref{sec:archit} compares some typical \ac{M2M}/\ac{IoT} protocol stacks. Section \ref{sec:scenarioSescription} describes the application scenario under consideration and Section \ref{sec:performanceEvaluation} shows some preliminary numerical results obtained via extensive simulation runs. Finally, the conclusions are provided in Section \ref{sec:conclusion}.

\section{Related Works}
\label{sec:rlatedWorks}
The work in \cite{related_work_1} considers \ac{M2M} terminals that communicate with a remote receiver via a satellite link. Each terminal transmits a fixed amount of data:  
\ac{RA} is used to deliver the first few messages, then the \ac{NCC} allocates reserved timeslots to each \ac{RCST}, in order to ensure a successfully delivery of data. The authors assess the system performance by setting the burst duration and by using three different reception modes at the receiver: \ac{TDMA}, \ac{FDMA}, \ac{PDMA} and \ac{TCDMA}. The throughput and the packet error rate are the performance metrics under consideration. The authors show that \ac{TCDMA} with multiuser detection outperforms both \ac{FDMA} and \ac{TDMA} in terms of throughput, thus minimizing the time required to serve a given set of \ac{M2M} terminals.
In \cite{related_work_2}, a satellite-based \ac{WSN} is considered. In order to handle a potentially large number of \ac{M2M} devices, the authors suggest to organize the nodes in clusters of different sizes. Within a cluster, the nodes communicate with the cluster-head (CH), which in turn forwards the collected data to the satellite gateway. The clustering mechanism aims at organizing the clusters in such a way that the application requirements (e.g., minimum required \ac{SNR}) can be met. Moreover, in order to cope with the rain fading that can affect the signal propagation, multiple interconnected satellite gateways are considered. Physical layer metrics are used to assess the performance level: \ac{SNR} and energy consumption. 
The aforementioned works 
do not consider any interactions with higher layer protocols, focusing on \ac{MAC} and physical layer metrics. Nonetheless, they consider \ac{M2M} scenarios involving satellite communications. 

Conversely, application-layer protocols are explicitly compared in \cite{related_work_4, related_work_3, related_work_5}. 
In \cite{related_work_4}, three classes of protocols for \ac{M2M} communications are compared: protocols targeting the \ac{SOA}; protocols implementing the \ac{REST} paradigm; 
and message-oriented protocols. 
As \ac{SOA} protocol, the authors consider OPC\footnote{A description of the OPC standard is available at https://opcfoundation.org/about/what-is-opc/} Unified Automation (UA), 
a platform-independent middleware, whereas \ac{CoAP} and MQTT are chosen as representative of \ac{REST} architectures and message-oriented protocols, respectively. The application scenario is represented by a cellular network delivering \ac{M2M} data, where reliability and real-time data exchanges are required. The completion time in a emulated cellular network is used as key performance indicator, and, according to the authors, OPC-based communications outperform \ac{CoAP} and MQTT-based data transfer, at the price of a larger overhead.  
MQTT is \ac{TCP}-based, and the contributions provided in \cite{Celandroni2016522, Gotta2014147, Bacco2014405, bacco2016m2m} shed some lights on how \ac{TCP} behaves in presence of a random access satellite link dominated by collisions, because it may represent a limiting factor on the achievable throughput in satellite environments. On the other side, \ac{CoAP} is \ac{UDP}-based, and a comparison is on order when dealing with random access satellite links, in order to provide some reference figures on the achievable performance level. 
A preliminary comparison between MQTT and \ac{CoAP} is provided in \cite{related_work_3}, in terms of bandwidth usage and latency. In order to compare the two protocols, the authors consider a simple scenario composed by a single MQTT publisher that sends data to a MQTT broker; similarly, the \ac{CoAP}-based scenario considers data exchanges between a \ac{HTTP} client and a \ac{CoAP} server. The metrics under consideration are the total amount of generated traffic and the average \ac{RTT}. Both reliable and non-reliable data transmissions are considered. In both cases, when no losses occur, \ac{CoAP} transfers less data and exhibits a shorter \ac{RTT} than MQTT, on average.
In \cite{related_work_5}, a satellite-based architecture is considered: multiple \ac{M2M} devices send data to a remote gateway. The performance provided by the use of MQTT and of \ac{CoAP} is evaluated, and the authors show that \ac{CoAP} can be properly tuned, in order to outperform MQTT even in the presence of high offered traffic. 

\section{Typical \ac{IoT} protocol stacks}
\label{sec:archit}
In \cite{igor}, the authors propose a general satellite network architecture for \ac{IoT}/\ac{M2M} application scenarios, where the use of satellite communications could provide some benefits, such as broadcast communications, large coverage also in suburban and rural areas, support for highly mobile nodes in absence of the fixed infrastructure. 
When comparing possible \ac{IoT} architectures, several different protocol stacks can be used, each providing different advantages. Two communication paradigms are typically considered in \ac{IoT} scenarios: request/response and \ac{PUB/SUB} ones.
\ac{IoT} nodes collect or produce new data in a event-driven or a time-driven fashion, typically. While the latter may exhibit regularity over time, the former shows variable traffic patterns. If the request/response paradigm is taken into account, the clients should periodically query the servers in order to retrieve fresh data. On high-delay links, like in the case of the satellites, the time needed to successfully complete data exchanges should be carefully evaluated. 
The \ac{PUB/SUB} paradigm, a possible alternative to the request/response one, allows the data consumers, or \textit{subscribers}, to receive any fresh data as soon as they are available at the data producers, or \textit{publishers}. A key feature of the \ac{PUB/SUB} paradigm is the decoupling between data producers and data consumers at an intermediate entity, called \textit{broker}. 
In topic-based \ac{PUB/SUB} systems, each data piece belongs to one or more \textit{topics}, or logical channels. The publishers send new data to the broker, specifying the topic(s) the data belong to. The broker keeps the list of the active topics and, for each topic, the list of the active subscribers. 
Each subscriber, in fact, declares its interests to the broker through an initial \textit{registration} procedure. After that, the mechanism is straightforward: the publishers send new data to the broker, which forward them to the subscribed nodes. On high-delay links, relying on a \ac{PUB/SUB}-based data exchange allows halving the delivery delay than relying on a request/response-based one.

\begin{figure}
	\centering
	\includegraphics[scale=0.95, clip=true, trim=0 0 0 0]{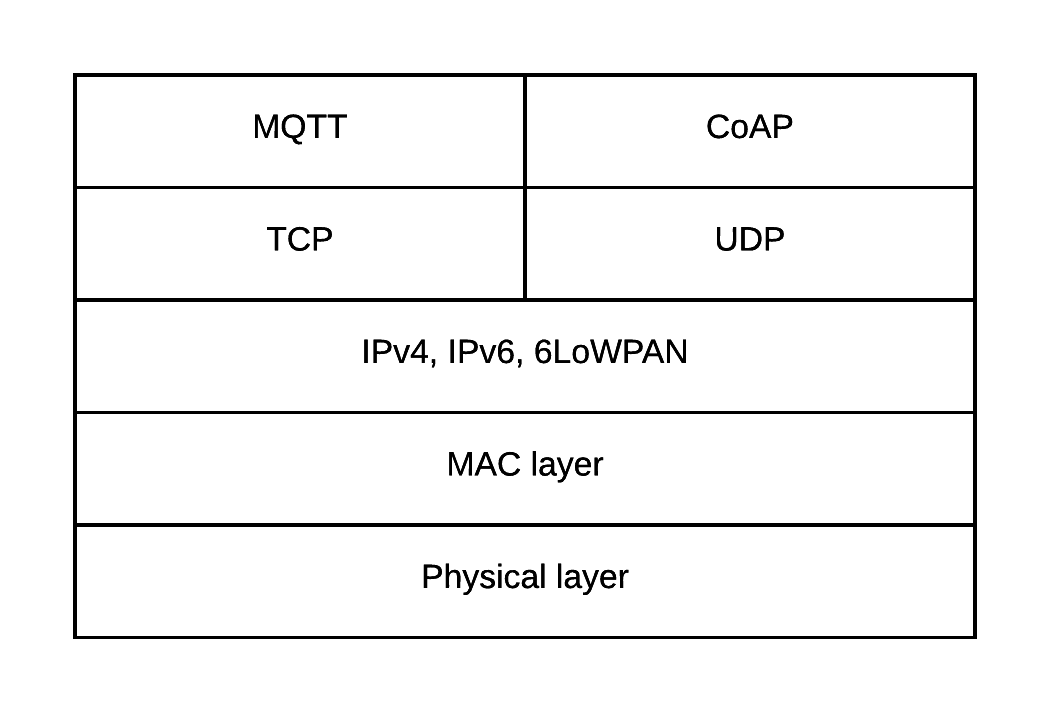}  	  \caption{Typical \ac{IoT} protocol stacks}
	\label{fig:stacks}
\end{figure}

\ac{CoAP} and MQTT are notable examples of application protocols implementing the aforementioned two paradigms: request/response the former, \ac{PUB/SUB} the latter. MQTT and \ac{CoAP} protocol stacks can be seen in Figure \ref{fig:stacks} and are discussed in Section \ref{subsec:mqtt} and \ref{subsec:coap}, respectively.

\subsection{MQTT protocol}
\label{subsec:mqtt}
MQTT is a \ac{M2M}/\ac{IoT} application protocol designed by IBM in 1999 for use in satellite networks. Since then, its use has largely widespread to terrestrial communications.
A typical MQTT data packet is composed of a 2 bytes long fixed header part, a variable header part whose size depends on the packet type, and a variable length payload. Each data packet is sent to the broker, which maintains the list of the active subscriptions and of the active topics.
Although reliable data transmission are inherently guaranteed by \ac{TCP}, MQTT offers three \ac{QoS} levels to deliver the messages. In fact, \ac{TCP} guarantees the reliability for the messages exchanged over the network connection between broker-publisher and broker-subscriber, but an \ac{E2E} mechanism is absent. To address that, MQTT provides additional reliability levels at the application.

\subsection{\ac{CoAP} protocol}
\label{subsec:coap}
\ac{CoAP} follows a \ac{REST} architectural style and is designed for resource-constrained environments. Each \ac{CoAP} server logically encapsulates a \textit{resource}, uniquely identify by a \ac{URI}. A \ac{CoAP} client sends a request by means of a \textit{Confirmable} or \textit{Non-confirmable} message, in order to retrieve the resource representation available at the server. If a \textit{Confirmable} message type is sent, an \ac{ACK} is expected to confirm the correct reception of data at the intended receiver; otherwise, an unreliable data exchange occurs (\textit{Non-confirmable} message type).
A \ac{CoAP} request contains the \ac{URI} of the resource and is typically performed by means of the HTTP \textit{GET} verb. The typical message format includes a fixed-size header (4 bytes), a variable-length \textit{Token} field (0-8 bytes), an \textit{options} field, and the payload. 
\ac{CoAP} is \ac{UDP}-based, and it provides optional reliability at the application layer. A $NSTART$-long [packets] transmission window\footnote{RFC 7252 defines $NSTART$ as the number of \textit{simultaneous outstanding interactions} (as in Section 4.7). For the sake of brevity, we refer to it as \textit{transmission window}.} is dictated by the specifications; in the default configuration, $NSTART$ is equal to one. If \textit{Confirmable} messages are sent, then the \ac{ARQ} mechanism in use is a simple Stop-and-Wait protocol,  
employing exponential back-off. This choice is motivated by the fact that the protocol is intended for low-power resource-constrained devices, where the implementation of more complex mechanisms can present some computational or technological issues because of the limited available resources. Anyway, \ac{CoAP} is a promising solution as application layer protocol, and its specifications open to the implementation and to the use of a different \ac{ARQ} mechanism, as the use of $NSTART > 1$ would require\footnote{Section 4.7 of RFC 7252 opens to the possibility of using $NSTART > 1$, if a congestion control mechanism is available.}. This would allow to take advantage of a large class of devices supporting more complex mechanisms and benefiting of a larger transmission window. In order to reduce the delivery delay of a request-response pattern, like in the case of \ac{CoAP}, the mechanisms described in the next two paragraphs can be applied.
\subsubsection{Observer pattern}
\ac{CoAP} specifications open to the implementation of the so-called \textit{observer} pattern,  
which provides a data exchange model semantically close to the \ac{PUB/SUB} one. A \ac{CoAP} client performs a registration to the server(s), indicating the \acp{URI} it is interested into. Anyway, there is still a difference with respect to the \ac{PUB/SUB} paradigm: the decoupling between data consumers and producers guaranteed by the publish/subscribe paradigm cannot be provided by the observer pattern. The absence of a intermediate entity leaves the server(s) in charge of keeping a list of the interested clients. The next paragraph explains how the use of a proxy can solve the aforementioned issue.
\subsubsection{\ac{CoAP} proxying}
In order to have a \ac{CoAP}-based configuration really similar to a \ac{PUB/SUB} one, a further step is necessary, which exploits the use of the \textit{proxying} functionality, as described in RFC 7252. 
A proxy is defined as a \ac{CoAP} endpoint that can be delegated by clients to perform requests on their behalf. Thus, a \ac{CoAP} proxy is an intermediate entity, which can actually decouple the clients from the servers. 

By implementing both the proxying functionality and the observer pattern, as we propose in this work, \ac{CoAP} behaves similarly to MQTT.
\subsection{Transport protocols}
A key difference between \ac{CoAP} and MQTT is the transport protocol they rely onto. MQTT is \ac{TCP}-based, which is connection oriented, thus a \ac{3WHS} procedure is needed to establish a connection. On the other side, \ac{CoAP} is \ac{UDP}-based\footnote{A \ac{TCP}-based \ac{CoAP} version is in a draft IETF proposal available at\\https://datatracker.ietf.org/doc/draft-ietf-core-coap-tcp-tls}, which realizes a connectionless communication and does not provide any congestion or flow control algorithms. While the use of \ac{TCP} can be of interest in some \ac{M2M}/\ac{IoT} scenarios \cite{bacco2016m2m, bacco2015tcp}, the majority of them would largely benefit of a lightweight transport protocol. 

In the following section, the performance provided by the use of MQTT and of \ac{CoAP} are compared, when the latter implements the observer pattern and the proxying functionality.


\section{Scenario description}
\label{sec:scenarioSescription}
\begin{figure}
	\centering
	\includegraphics[scale=0.35]{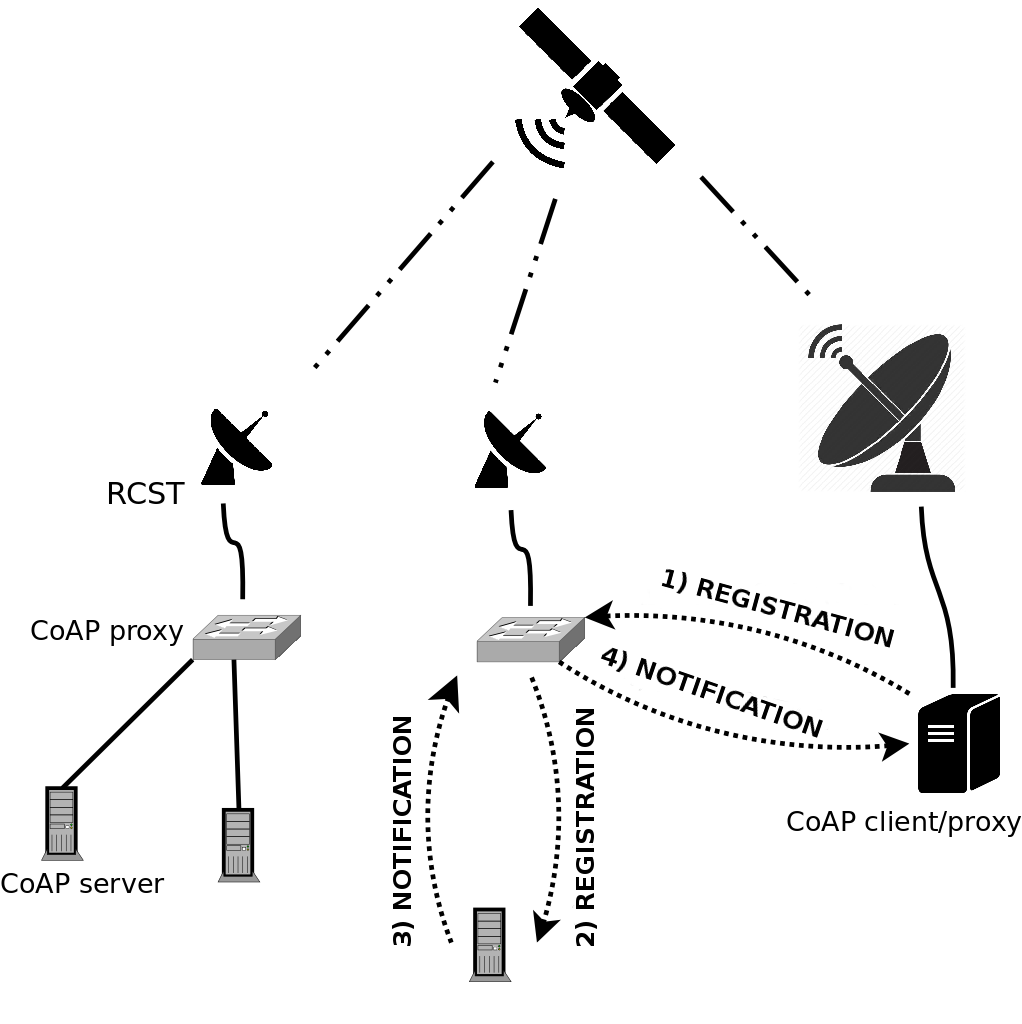}  
	\caption{The \ac{CoAP}-based scenario under consideration. The dotted lines show a typical setup procedure at application layer (steps 1 and 2) and the notification of new messages via proxy (steps 3 and 4).}
	\label{fig:coap_scenario}
\end{figure}

In this section, the scenario under consideration is described, and the logical architecture is visible in Figure \ref{fig:coap_scenario}. 
Several \ac{CoAP} servers produce data that are sent to the \ac{CoAP} proxies. Each proxy is in charge of delivering the received data to a remote \ac{CoAP} client via \ac{DVB-RCS2}-compliant \acp{RCST}. 
Thus, we assume that the \ac{CoAP} servers, which encapsulate available resources, act as data producers. For instance, such an architecture can be applied to low-altitude \ac{UAV} swarms, whose use is growing more and more common in several application fields, such as the case of precision agriculture \cite{bacco2014uavs}. A master \ac{UAV} acts as a proxy, collecting data from other \acp{UAV} in the same swarm and delivering data via satellite to a remote data center.

We recall that \ac{CoAP} is \ac{UDP}-based, thus the \ac{ARQ} protocol must be implemented at the application layer, if reliable delivery is expected. In Section \ref{subsec:modCOAP}, the \ac{CoAP} settings in use are described, while Section \ref{subsec:sc} deepens the description of the system configuration.

\subsection{\ac{CoAP} protocol implementation}
\label{subsec:modCOAP}
We implemented the \ac{CoAP} protocol as a Network Simulator 3 (NS-3) 
module, along with the observer pattern and the proxying functionality. The numerical results in Section \ref{sec:performanceEvaluation} are based on the use of those extensions, thus the typical \ac{CoAP} request/response paradigm is substituted by a \ac{PUB/SUB}-like mechanism. The reason behind the latter choice is straightforward: removing the need for a request, the data delivery delay is reduced, because fresh data are available to a client as soon as they are generated or collected by a server. On high-delay links, a \textit{push} strategy can provide large gains with respect to a \textit{pull} one, for instance in terms of delivery delay.



\subsection{System configuration}
\label{subsec:sc}
We consider a network composed by a \ac{GEO} satellite and 
a large number of \ac{CoAP} servers, connected to \ac{CoAP} proxies; each proxy is connected to a \ac{RCST} (see Figure \ref{fig:coap_scenario}).
It is worth to underline here that a single proxy (instead of multiple ones) can be used if a single network is desired; multiple proxies are to be used if separated networks are required. In Figure \ref{fig:coap_scenario}, we refer to the more general case with multiple separated networks.
\ac{CoAP} servers produce \ac{M2M}/\ac{IoT}-like data that is delivered to a remote \ac{CoAP} client. If more clients were present on the remote side, a connection per client would be opened, thus increasing the contention level on the random access channel. Alternatively, a receiving proxy can be placed on the remote side, too, in order to have a single connection per sender proxy to the receiving one. Thus, the scenario under consideration can be considered as representative of both aforementioned cases, because the \ac{CoAP} client in Figure \ref{fig:coap_scenario} can be substituted by a \ac{CoAP} proxy, then connected to multiple clients.

In the extensive simulations we ran, data sources begin the transmission according to an exponentially distributed inter-arrival time with parameter $\lambda$. The data payload length is randomly drawn, with probability $1/i$, from $i$ Pareto distributions with parameters $x_m^i > 0$ and $\alpha^i > 0$, where $i \in \{1,2,3\}$. 
The three distributions are here meant to represent small, medium and large application \ac{M2M}/\ac{IoT} payload lengths, as generated or collected by the server(s). More technically, each \ac{CoAP} server sends a burst of packets, then forwarded by the proxy, 
which are packed into a bulk of \ac{DVB-RCS2} RA blocks, according to the specifications of Waveform 14\footnote{The waveform specifications are drawn from Table A-1 \textit{"Reference Waveforms for Linear Modulation Bursts"} in \cite{dvbrcs2}.}, reported in Table \ref{table:settings}. A single timeslot can be used per \ac{RA} block by each \ac{RCST}, according to typical configurations of \ac{DVB-RCS2} systems. The \ac{MAC} protocol in use is \ac{CRDSA} \cite{crdsa}, configured with 3 replicas. Time is slotted and each \ac{RA} block is composed of 64 timeslots. The \ac{MAC} queue length is set to an arbitrary large value and both return and forward links are assumed to be error-free. In the return link, collisions can occur, thus retransmissions are triggered in order to ensure a reliable data delivery. \acp{ACK} are assumed to be always correctly received.

\section{Performance Evaluation}
\label{sec:performanceEvaluation}
\begin{figure}
	\centering
	\includegraphics[scale=0.38]{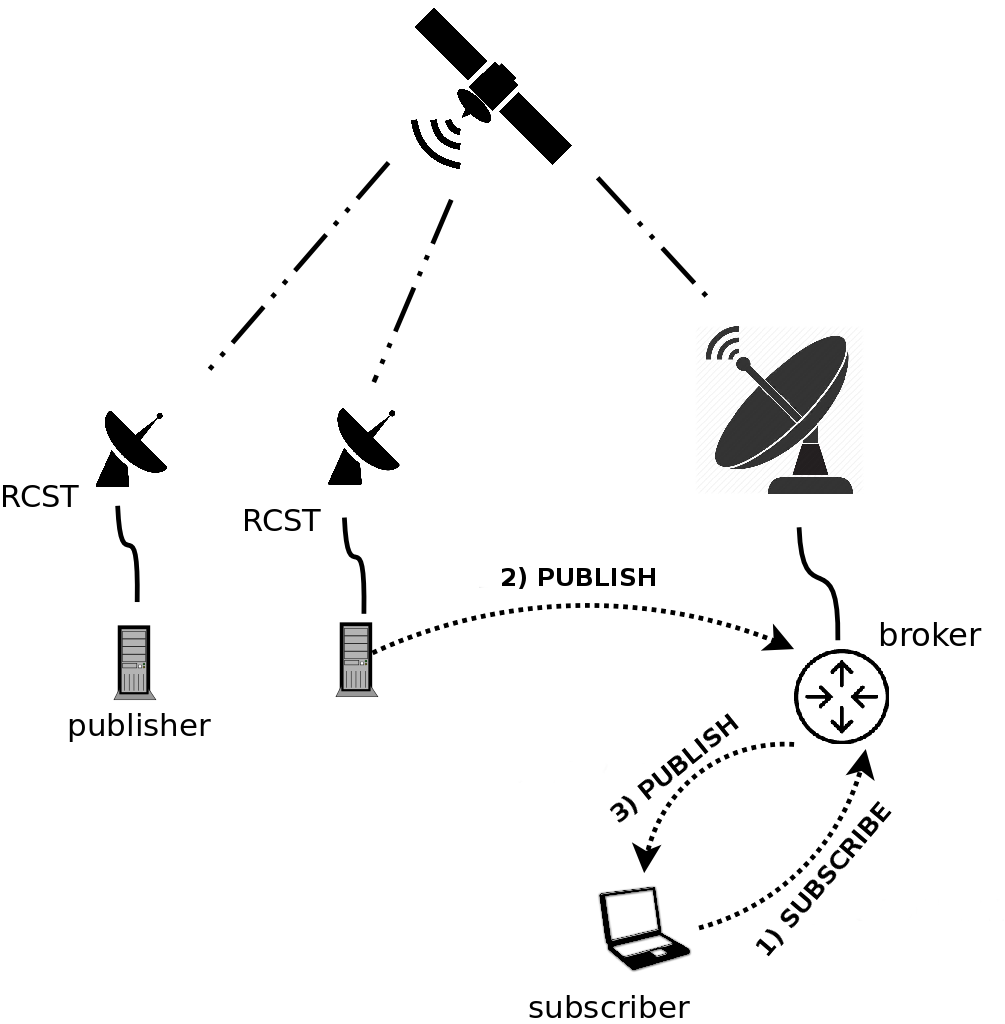}  
	\caption{The MQTT-based scenario in use for comparison. The dotted lines show a typical setup procedure at application layer (step 1) and the notification of new messages via broker (steps 2 and 3).}
	\label{fig:mqtt_scenario}
\end{figure}

The scenario presented in the previous section is here numerically evaluated and then compared to the MQTT-based scenario in \cite{advances}, which is sketched in Figure \ref{fig:mqtt_scenario}.
In this work, the length of the transmission windows of the \ac{CoAP} servers ranges into $NSTART \in [1,100]$. Thanks to this, we explore the possibility to reduce the completion time by increasing $NSTART$. 
In the following, a Go-Back-N \cite{burton1972errors} \ac{ARQ} protocol, employing exponential backoff, is in use if $NSTART > 1$.
The following numerical results are based on extensive simulation runs based on the use of S-NS3 \cite{hytonen2014satellite}, 
a satellite network extension to NS-3 platform. The simulator parameters have values as reported in Table \ref{table:settings} and a minimum of 1000 data exchanges per scenario has been simulated, in order to ensure statistical reliability.
\begin{table}
	\begin{center}
		\begin{tabular}{|c|c|} 							\hline
			\textbf{Name} & \textbf{Value} 				\\ \hline
			\hline
			\ac{RA} scheme & 3-\ac{CRDSA}					\\ \hline 
			\ac{RA} blocks per superframe & 1				\\ \hline
			\ac{RA} block duration & 13 [ms]				\\ \hline
			Timeslots per \ac{RA} block & 64				\\ \hline
			Gross slot size & 188 [B]						\\ \hline
			Net slot size & 182 [B]						\\ \hline
			Bandwidth & 8012820 [Hz]						\\ \hline
			Roll off & 0.2								\\ \hline  
			Carrier spacing & 0.3	[Hz]					\\ \hline
			Nominal \ac{RTT} & 0.52 [s]					\\ \hline
			Pareto distributions &$x_m^1 = 931$ [B] 	\\ 
			&$x_m^2 = $ 9532 [B] 						\\ 
			&$x_m^3 = $ 47663 [B] 						\\
			&$\alpha^1 = \alpha^2 = \alpha^3 = 1.1$ 		\\ \hline
		\end{tabular}
	\end{center}
	\caption{Simulator setup parameters}
	\label{table:settings}
\end{table}

\begin{table}
	\begin{center}
		\begin{tabular}{|c|c|} 							\hline
			\textbf{Protocol stack} & \textbf{Avg. aggregated goodput}	\\ \hline
			\hline
			\ac{CoAP}/UDP ($NSTART = 1$) 	& 32.14 [KB/s]			\\ \hline
			\ac{CoAP}/UDP ($NSTART = 2$) 	& 40.46 [KB/s]			\\ \hline
			\ac{CoAP}/UDP ($NSTART = 3$) 	& 43.21 [KB/s]			\\ \hline
			\ac{CoAP}/UDP ($NSTART = 4$) 	& 45.58 [KB/s]			\\ \hline
			\ac{CoAP}/UDP ($NSTART = 5$) 	& 50.1 [KB/s]			\\ \hline
			\ac{CoAP}/UDP ($NSTART = 10$) 	& 57.1 [KB/s]			\\ \hline
			\ac{CoAP}/UDP ($NSTART = 100$) 	& 77.5 [KB/s]			\\ \hline \hline
			MQTT/\ac{TCP} 					& 50.6 [KB/s]			\\ \hline
		\end{tabular}
	\end{center}
	\caption{Average aggregated goodput at MQTT broker or at \ac{CoAP} client/proxy in the scenarios under consideration}
	\label{table:goodput}
\end{table}

\begin{figure}
	\centering
	\includegraphics[scale=0.4, clip=true, trim=40 50 0 0]{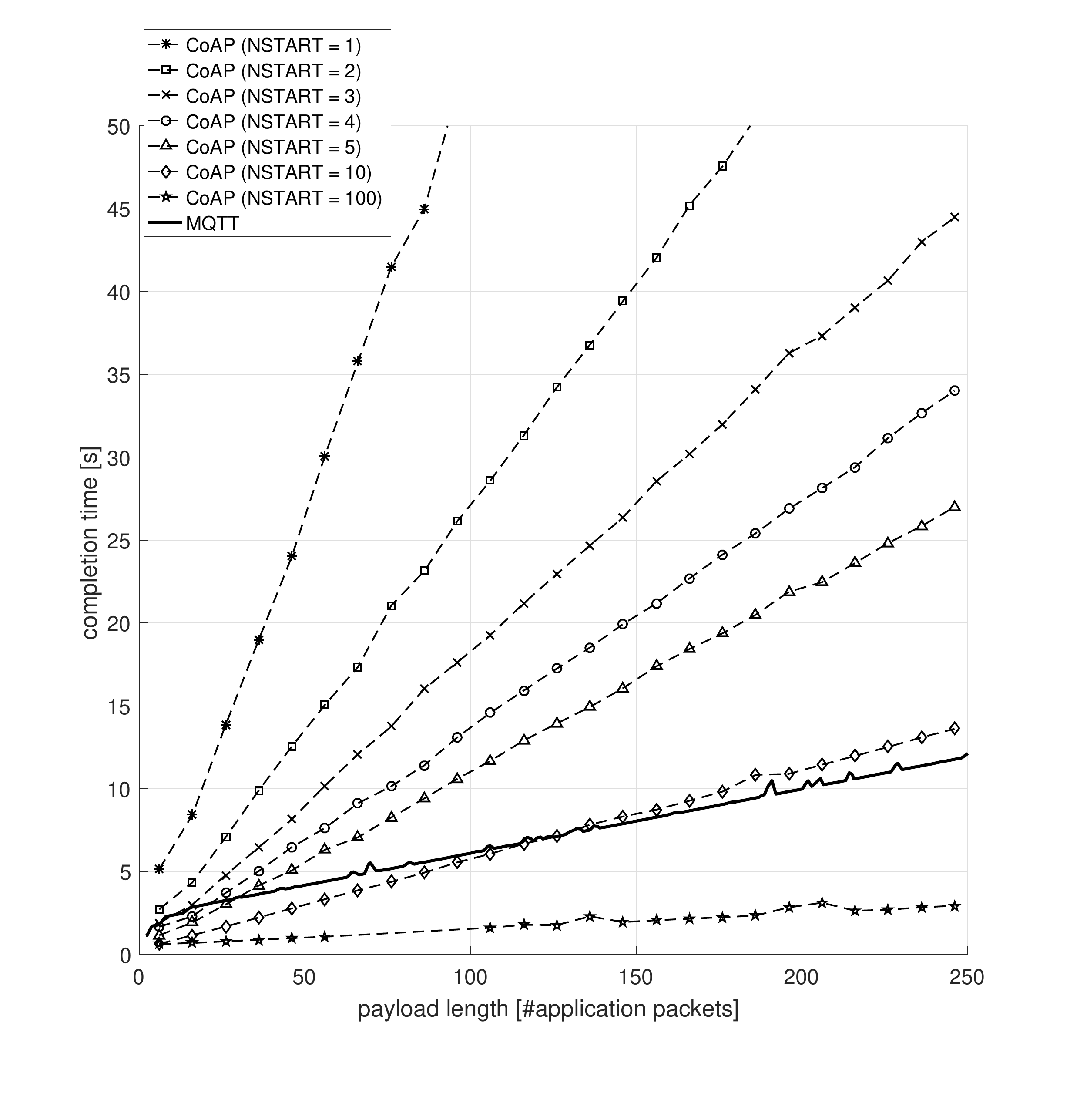}
	\caption{Completion time of MQTT and \ac{CoAP} data exchanges per \ac{CoAP} proxy/client or MQTT publisher.}
	\label{fig:CT}
\end{figure}

In Figure \ref{fig:CT}, the completion time of \ac{CoAP} and MQTT data exchanges is visible, in presence of low/moderate load. The completion time is plotted against an increasing number of application packets sent per data exchange, as readable on the x-axis. 
Seven increasing $NSTART$ values have been selected in the \ac{CoAP}-based scenario: the use of a larger value provides a lower completion time, as expected. Anyway, the latter is valid only in presence of a low/medium traffic profile on the \ac{RA} channel, so that erasures due to collisions are unlikely to occur. Table \ref{table:results_IAT_1s} reports the normalized \ac{MAC} offered load for each $NSTART$ value and its 25th and 75th percentiles, when $\lambda^{-1} = 1$ [s]. The collision rate is almost negligible for the load intervals under consideration. 
A low/medium traffic profile is used, in order to avoid congestion phenomena.
\begin{table}
	\begin{center}
		\begin{tabular}{|c|c|c|c|} \hline
			\textbf{NSTART} &\multicolumn{3}{|c|}{\textbf{Normalized \ac{MAC} offered load}} 
			\\ \hline
			& \textbf{mean} & \textbf{25th p.} & \textbf{75th p.} \\ \hline
			1 & 0,0671 & 0,031 & 0,09 
			\\ \hline
			2 &0,0820 & 0,047 & 0,11 
			\\ \hline
			3 &0,0880 & 0,047 & 0,12 
			\\ \hline
			4 &0,0929 & 0,048 & 0,12 
			\\ \hline
			5 &0,1016 & 0,048 & 0,14 
			\\ \hline
			10 &0,1161	& 0,062 & 0,15	
			\\ \hline
			100 &0,1172 & 0,063 & 0,16 
			\\ \hline  
		\end{tabular}
	\end{center}
	\caption{Normalized \ac{MAC} offered load for increasing $NSTART$ values if \ac{CoAP} is in use at the application layer} 
	\label{table:results_IAT_1s}
\end{table}

The default \ac{CoAP} configuration ($NSTART=1$) provides a quite large completion time, even for small amounts of data, sub-utilizing the available system resources. In case of larger values, the completion time decreases and, for $NSTART=100$, \ac{CoAP} provides a lower completion time than MQTT. It is worth underlining here that the \ac{BDP} of the satellite link is $\approx 40$ \ac{CoAP} packets; thus, when $NSTART = 100$ and the burst length is larger than \ac{BDP}, backlogging is present.

The completion time of the MQTT-based scenario depends on \ac{TCP}\footnote{TCP NewReno (RFC 6582) is in use in this scenario.} congestion control algorithm; as the \ac{TCP} congestion window increases over time, the curve exhibits an almost linear trend to a first approximation, in presence of a low collision rate on \ac{RA} channels. 

If looking at the completion time, a comparable value is obtained with a \ac{CoAP} configuration with $NSTART=10$. Anyway, the different trends in the completion time provided by MQTT and \ac{CoAP} are clearly visible: in the first part, the completion time in MQTT-based scenario increases faster than in the \ac{CoAP}-based scenario because of the small \ac{TCP} congestion window. As soon as \ac{TCP} congestion window increases because of larger payload lengths, the completion time reduces accordingly. 

Table \ref{table:goodput} shows the average aggregated goodput at \ac{CoAP} client/proxy (see Figure \ref{fig:coap_scenario}), compared with the same at MQTT broker (see Figure \ref{fig:mqtt_scenario}). Thanks to the lower overhead provided by the \ac{CoAP}/\ac{UDP} stack, it outperforms the MQTT/TCP stack. In fact, even if MQTT/\ac{TCP} is approximately equivalent to using \ac{CoAP}/\ac{UDP} with $NSTART=10$, the larger overhead reduces the achievable goodput w.r.t. the latter configuration, as shown in Table \ref{table:goodput}.
Eventually, some considerations are in order: in \ac{IoT}/\ac{M2M} scenarios, the use of \ac{CoAP} can provide some advantages over MQTT, because the length of the transmission window can be set at the application level, as well as the \ac{ARQ} algorithm in use, thus providing a larger flexibility. Furthermore, the \ac{CoAP}-based protocol stack exhibits a lower overhead, which is desirable in such application scenarios. 

\section{Conclusions}
\label{sec:conclusion}
This work focuses on a comparison between two of the largely used \ac{IoT}/\ac{M2M} protocol stacks, based on the use of \ac{CoAP} and MQTT protocols, implementing the request/response and the \ac{PUB/SUB} communications paradigm, respectively. The \ac{PUB/SUB} paradigm can bring large benefits in satellite-based architectures, because of the reduction of the delivery time thanks to the fact that fresh data are sent to registered subscribers as soon they are produced. Because of the latter, in this work, we investigated the use of the \ac{CoAP} protocol in conjunction with the so-called \textit{observer} pattern and the \textit{proxying} functionality, in order to exploit the advantages provided by the \ac{PUB/SUB} paradigm, which also provides a fairer comparison than relying on the default \ac{CoAP} implementation. 
A qualitative comparison is provided in this work, together with some preliminary numerical results, highlighting how the performance level provided by the use of \ac{CoAP} outperforms the one provided by MQTT on \ac{RA} satellite channels, and underlining the flexibility that easily tunable settings at application layer provide w.r.t to lower layer settings.
Future work will focus on \ac{M2M}/\ac{IoT} communications in presence of higher traffic rates, where the performance provided by \ac{CoAP} may still need further investigation.

\section*{Acknowledgments}
This work has been partially supported by the Tuscany region in the framework of SCIADRO project (FAR-FAS 2014), and by SatNEx (Satellite Network of Experts) programme, IV phase.

\IEEEtriggeratref{0}
\balance
\bibliographystyle{IEEEtran}
\bibliography{references}

\end{document}

%% file: acr.tex
\acrodef{PER}{Packet Erasure Rate}
\acrodef{PLR}{Packet Loss Rate}
\acrodef{PRR}{Packet Received Ratio}
\acrodef{BER}{Bit Error Rate}
\acrodef{RTT}{Round-Trip Time}
\acrodef{TCP}{Transmission Control Protocol}
\acrodef{UDP}{User Datagram Protocol}
\acrodef{AIMD}{Additive Increase Multiplicative Decrease}
\acrodef{PEP}{Performance Enhancing Proxy}
\acrodef{cwnd}{congestion window}
\acrodef{STP}{Satellite Transport Protocol}
\acrodef{BDP}{Bandwidth-Delay Product}
\acrodef{QoS}{Quality of Service}
\acrodef{LoS}{Line of Sight}
\acrodef{NLoS}{Non Line of Sight}
\acrodef{BLoS}{Beyond Line of Sight}
\acrodef{QoE}{Quality of Experience}
\acrodef{ACM}{Adaptive Coding and Modulation}
\acrodef{DAMA}{Demand Assignment Multiple Access}
\acrodef{VANET}{Vehicular Ad-Hoc Network}
\acrodef{RA}{Random Access}
\acrodef{CRA}{Contention Resolution ALOHA}
\acrodef{SA}{Slotted ALOHA}
\acrodef{DSA}{Diversity Slotted ALOHA}
\acrodef{OBU}{On-Board Unit}
\acrodef{CRDSA}{Contention Resolution Diversity Slotted ALOHA}
\acrodef{SIC}{Successive Interference Cancellation}
\acrodef{ARQ}{Automatic Repeat reQuest}
\acrodef{SC-ARQ}{Selective-Coded ARQ}
\acrodef{SR-ARQ}{Selective-Repeat ARQ}
\acrodef{IRSA}{Irregular Repetition Slotted ALOHA}
\acrodef{CGC}{Complementary Ground Component}
\acrodef{RSU}{Road Side Unit}
\acrodef{ACK}{Acknowledgment}
\acrodef{NACK}{Negative Acknowledgment}
\acrodef{DVB-SH}{Digital Video Broadcasting - Satellite Services to Handhelds}
\acrodef{DVB-H}{Digital Video Broadcasting - Handheld}
\acrodef{DVB-RCS2}{Digital Video Broadcasting - Return Channel via Satellite, II generation}
\acrodef{SACK}{Selective Acknowledgment}
\acrodef{SNACK}{Selective Negative Acknowledgment}\acrodef{SNACK}{Selective Negative Acknowledgment}
\acrodef{SNIR}{Signal to Noise plus Interference Ratio}
\acrodef{SCPS-TP}{Space Communications Protocol Specifications - Transport Protocol}
\acrodef{CCSDS}{Consultative Committee for Space Data Systems}
\acrodef{ESA}{European Space Agency}
\acrodef{NASA}{National Aeronautics and Space Administration}
\acrodef{BSM}{Broadband Satellite Multimedia}
\acrodef{RLNC}{Random Linear Network Coding}
\acrodef{NC}{Network Coding}
\acrodef{FIFO}{First In, First Out}
\acrodef{FCFS}{First Come, First Served}
\acrodef{BLER}{Block Error Rate}
\acrodef{GEO}{Geosynchronous}
\acrodef{LEO}{Low Earth Orbit}
\acrodef{FTP}{File Transfer Protocol}
\acrodef{CRC}{Cyclic Redundancy Check}
\acrodef{MAC}{Media Access Control}
\acrodef{HTTP}{Hypertext Transfer Protocol}
\acrodef{ISP}{Internet Service Provider}
\acrodef{MSS}{Maximum Segment Size}
\acrodef{BIC}{Binary Increase Congestion control}
\acrodef{AQM}{Active Queue Management}
\acrodef{XCP}{eXplicit Control Protocol}
\acrodef{ECN}{Explicit Congestion Notification}
\acrodef{CA}{Congestion Avoidance}
\acrodef{RED}{Random Early Detection}
\acrodef{TD}{Triple-Duplicate}
\acrodef{TO}{Time Out}
\acrodef{IP}{Internet Protocol}
\acrodef{WMN}{Wireless Mesh Network}
\acrodef{ssthresh}{Slow-Start threshold}
\acrodef{MPE-IFEC}{Multi Protocol Encapsulation - Inter-burst Forward Error Correction}
\acrodef{FEC}{Forward Error Correction}
\acrodef{ML}{Maximum Likelihood}
\acrodef{CRC}{Cyclic Redundancy Check}
\acrodef{P2P}{Peer-to-Peer}
\acrodef{FMT}{Fade Mitigation Technique}
\acrodef{SGD}{Smart Gateway Diversity}
\acrodef{NCC}{Network Control Centre}
\acrodef{ModCod}{Modulation and Coding}
\acrodef{FIFO}{First-In-First-Out}
\acrodef{WRR}{Weighted Round Robin}
\acrodef{WFQ}{Weighted Fair Queuing}
\acrodef{NS}{Network Simulator}
\acrodef{GSE}{Generic Stream Encapsulation}
\acrodef{PDF}{Probability Density Function}
\acrodef{CDF}{Cumulative Density Function}
\acrodef{CoV}{Coefficient of Variation}
\acrodef{MSC}{Message Sequence Chart}
\acrodef{ESA}{European Space Agency}
\acrodef{LIU}{Lebanese International University}
\acrodef{TUM}{Technical University of Munich}
\acrodef{MSCE-CS}{Master of Science in Communications Engineering - Communications Systems}
\acrodef{DLR}{German Aerospace Center}
\acrodef{NC-SGD}{Network Coding for SGD}
\acrodef{ATSP}{Advanced Transport Satellite Protocol}
\acrodef{STP}{Satellite Transport Protocol}
\acrodef{WMN}{Wireless Mesh Network}
\acrodef{SNR}{Signal-to-Noise Ratio}
\acrodef{SINR}{Signal-to-Interference-plus-Noise Ratio}
\acrodef{LMS}{Land Mobile Satellite}
\acrodef{LTE}{Long-Term Evolution}
\acrodef{M2M}{Machine-to-Machine}
\acrodef{IoT}{Internet of Things}
\acrodef{RA}{Random Access}
\acrodef{UAV}{Unmanned Aerial Vehicle}
\acrodef{UAS}{Unmanned Aerial System}
\acrodef{FANET}{Flying Ad-Hoc Network}
\acrodef{MANET}{Mobile Ad-Hoc Network}
\acrodef{VANET}{Vehicle Ad-Hoc Network}
\acrodef{C2}{Command and Control}
\acrodef{DTN}{Delay Tolerant Network}
\acrodef{COTS}{Commercial Off-the-Shelf}
\acrodef{IETF}{Internet Engineering Task Force}
\acrodef{CoAP}{Constrained Application Protocol}
\acrodef{URI}{Uniform Resource Identifier}
\acrodef{PUB/SUB}{Publish / Subscribe}
\acrodef{RCST}{Return Channel Satellite Terminal}
\acrodef{TDMA}{Time Division Multiple Access}
\acrodef{FDMA}{Frequency Division Multiple Access}
\acrodef{TCDMA}{Turbo Code Division Multiple Access}
\acrodef{PDMA}{Power Division Multiple Access}
\acrodef{WSN}{Wireless Sensor Network}
\acrodef{REST}{Representational State Transfer}
\acrodef{EDGE}{Enhanced Data rates for GSM Evolution}
\acrodef{UMTS}{Universal Mobile Telecommunications System}
\acrodef{LTE}{Long-Term Evolution}
\acrodef{E2E}{End-to-End}
\acrodef{3WHS}{Three-way Handshake}
\acrodef{SCADA}{Supervisory Control And Data Acquisition}
\acrodef{SOA}{Service-Oriented Architecture}

\acrodefplural{RSU}[RSUs]{Road Side Units}
\acrodefplural{VANET}[VANETs]{Vehicular Ad-Hoc Networks}
\acrodefplural{OBU}[OBUs]{On-Board Units}
\acrodefplural{UAV}[UAVs]{Unmanned Aerial Vehicles}
\acrodefplural{FANET}[FANETs]{Flying Ad-Hoc Networks}
\acrodefplural{MANET}[MANETs]{Mobile Ad-Hoc Networks}
\acrodefplural{VANET}[VANETs]{Vehicle Ad-Hoc Networks}
\acrodefplural{UAS}[UASs]{Unmanned Aerial Systems}
\acrodefplural{RCST}[RCSTs]{Return Channel Satellite Terminal}